\documentclass[12pt]{article}
\usepackage{amsmath}
\usepackage{amsfonts}
\usepackage{amssymb}
\usepackage{xcolor}
\usepackage{color}
\usepackage[english]{babel}
\usepackage{graphics}
\usepackage{graphicx}
\begin{document}
\title{The true quantum face of the "exponential" decay law}
\author{K. Urbanowski\footnote{e--mail:  K.Urbanowski@if.uz.zgora.pl}, \\
\hfill\\
University of Zielona G\'{o}ra, Institute of Physics, \\
ul. Prof. Z. Szafrana 4a, 65--516 Zielona G\'{o}ra, Poland.
}
\maketitle


\begin{abstract}
Results of theoretical studies of the quantum unstable systems caused that
there are rather widespread belief that a universal feature od the quantum decay process is the presence of three
time regimes of the decay process:
the early time (initial) leading to the Quantum Zeno (or Anti Zeno) Effects, "exponential" (or "canonical") described by the decay law of the exponential form, and late time characterized by the decay law having inverse--power law form. Based on the fundamental principles of the quantum theory
we give the proof that there is no time interval in which the survival probability (decay law) could be a decreasing function of time of the purely exponential form
but even at the "exponential" regime the decay curve is oscillatory modulated
with a smaller or a large amplitude of oscillations depending on parameters of the model considered.
\end{abstract}

\section{Introduction}

The discovery of radioactivity in the nineteenth century initialized  the study of the process of decay of radioactive elements. Experiments have shown that the radioactive decay of the sample of radioactive elements is a process extended in time, and that with the passage of time  the number of elements in the sample, which emits radioactive radiation,
decreases. These observations and assumption that the decay rate follows the laws of probability led Rutherford and Sody to the formulation of radioactive decay law as a function of time \cite{rutheford,rutheford1}. This radioactive decay law allows to determine the number  $N (t)$ of atoms of the radioactive element at the instant $t$  knowing the initial number  $N_{0} = N(0)$ of them  at  initial instant of time $t_{0} =0$ and has the exponential form:  $N(t) =N_{0}\,\exp\,[-\lambda t]$, where $\lambda > 0$ is a constant.
Since then, the belief that the  decay law has the exponential form has become common.
The rise of Quantum Mechanics    led to an understanding that the radioactive decay similarly to the process of  emission of photons  by excited atoms are  time dependent quantum processes. So the question arose how to describe such  processes within the quantum theory. Probably the most known attempt to solve this problem is the Weisskopf--Wigner
theory of spontaneous emission \cite{WW}. Considering the excited atomic levels and applying the Shr\"{o}dinger equation to describe the time evolution Weisskopf and Wigner
found that to a good approximation the non--decay probability of the excited levels is a decreasing function of time having exponential form.
Further studies of the quantum decay processes showed that basic principles of the quantum theory  does not allow them to be described by an exponential decay law at very late times
\cite{khalfin,fonda} and at initial stage of the decay process (see e.g. \cite{fonda} and references therein). Theoretical analysis shows that at late times the survival probability (i. e. the decay law) should tends to zero as
$t \to \infty$ much more slowly than any exponential function of time and that
as a function of time
it has the inverse power--like form
at this  regime of time \cite{khalfin,fonda}.
There was many unsuccessful attempts to verify experimentally predicted deviations from the exponential form of the decay law at late times regime (see eg. \cite{norman}). The first experimental evidence of deviations of the decay law from exponential form at such a time regime was reported in \cite{rothe}. The early times properties of the decay process lead to the so called Quantum Zeno Effect \cite{misra,fischer}, that is to slowing down sufficiently frequently observed decay process up to stop it down in the case of the continuously observed the unstable system. The experimental confirmation of this effect was reported, e.g.  in \cite{itano} and recently in \cite{patil}.
All these results of theoretical and experimental researches caused that
there are rather widespread belief that a universal feature of the quantum decay process is the presence of three
time regimes of such a decay process:
the early time (initial), exponential (or "canonical"), and late time having inverse--power law form \cite{peshkin}.
This  belief is reinforced by a numerous presentations in the literature of decay curves obtained for quantum models of unstable systems. The typical form of such a decay curve one can find in Fig. (\ref{f1}).
In this context, each experimental  evidence of oscillating decay curve at times of the order of life time is considered as an anomaly caused by a new quantum effects or new interactions (see eg. \cite{litwinov1,kienle}).
The question arises, if indeed in the case of one component quantum unstable systems these oscillations of the decay process at the "exponential" regime are an anomaly, or perhaps universal feature of quantum decay processes.
Here
we give the proof that there is no time interval in which the survival probability (decay law) could be a decreasing function of time of the purely exponential form.
We also show that even in the case of a single component unstable system the decay  curve has an oscillatory form with a smaller or a large amplitude of oscillations depending on the model considered.

\section{Preliminaries}
The main information about properties of  quantum unstable   systems
is contained in their decay law, that is in their survival probability.
Let the reference frame ${\cal O}$ be the common inertial rest
frame for the observer and for the unstable system.
Then  if one knows that the system in the rest frame is in the initial unstable
state $|\phi\rangle \in {\cal H}$, (${\cal H}$ is
 the Hilbert space of states of the considered system), which was prepared at the initial instant $t_{0} =0$,
then one can calculate
its survival probability (the decay law), ${\cal P}(t)$, of the unstable state $|\phi\rangle$ decaying
in vacuum, which equals
\begin{equation}
{\cal P}(t) = |a(t)|^{2}, \label{P(t)}
\end{equation}
where $a(t)$ is  the probability amplitude of finding the system at the
time $t$ in the initial unstable state $|\phi\rangle$,
\begin{equation}
a(t) = \langle \phi|\phi (t) \rangle . \label{a(t)}
\end{equation}
and $|\phi (t)\rangle$ is the solution of the Schr\"{o}dinger equation
for the initial condition  $|\phi (0) \rangle = |\phi\rangle$:
 \begin{equation}
i\hbar \frac{\partial}{\partial t} |\phi (t) \rangle = H |\phi (t)\rangle.  \label{Schrod}
\end{equation}
 Here $|\phi \rangle, |\phi (t)\rangle \in {\cal H}$,  and
 $H$ denotes the total self--adjoint Hamiltonian for the system considered. Note that if $|\phi\rangle$ represents an unstable state then it cannot be
 an eigenvector for $H$: In such a case the eigenvalue equation $H|\phi\rangle = \epsilon_{\phi} |\phi\rangle$ has no solutions for $|\phi \rangle$ under considerations.

There is $|\phi (t)\rangle = U(t) |\phi\rangle$, where $U(t)$ is unitary evolution operator and $U(0) = \mathbb{I}$ is the unit operator. Thus
$a(t) \equiv \langle \phi|U(t)|\phi\rangle$.
The one--parameter family of unitary operators $U(t)$ forms group: $U(t_{1})\,U(t_{2}) = U(t_{1} + t_{2})$. The the total Hamiltonian $H$ of the system is a generator of this group.
This means that operators $H$ and $U(t)$ have common eigenfunctions.

From the results of theoretical studies of the problem which one can find in
the literature it is known that the
amplitude $a(t)$, and thus the decay law ${\cal P}(t)$ of the
unstable state $|\phi\rangle$, are completely determined by the
density of the energy distribution $\omega(E)$ for the system
in this state \cite{fock},
\begin{equation}
a(t) = \int_{Spec.(H)} \omega(E)\;e^{\textstyle{-\frac{i}{\hbar}\,E\,t}}\,dE, \label{a-spec-1}
\end{equation}
where $\omega(E) \geq 0$  and $a(0) = 1$,
(see also: \cite{khalfin,fonda,nowakowski,giraldi}).
From this
relation and from the Riemann--Lebesgue lemma it follows that
$|a(t)| \to 0$ as $t \to \infty$. It is because from the normalization
condition $a(0) = 1$
it follows that $\omega (E)$ is an absolutely integrable function.
These properties are the essence of the so-called Fock--Krylov theory of unstable states \cite{fock}.
(Note that this approach is also applicable in Quantum Field Theory models \cite{giacosa2,goldberger}).

Khalfin in  \cite{khalfin}
assuming that the spectrum of $H$ must be bounded
from below, $(Spec.(H) = [E_{min},\infty)$ and $E_{min} > -\infty)$, that is that
$\omega (E) = 0$ for $E < E_{min}$,
and using the Paley--Wiener
Theorem  \cite{Paley}
proved that in the case of unstable
states there must be
$|a(t)| \; \geq \; A\,\exp\,[- b \,t^{q}]$,
for $|t| \rightarrow \infty$. Here $A > 0,\,b> 0$ and $ 0 < q < 1$.
Therefore
the decay law ${\cal P}_{\phi}(t)$ of unstable
states decaying in the vacuum, (\ref{P(t)}), can not be described by
an exponential function of time $t$ if time $t$ is suitably long, $t
\rightarrow \infty$, and that for these lengths of time ${\cal
P}_{\phi}(t)$ tends to zero as $t \rightarrow \infty$  more slowly
than any exponential function of $t$.
As it was mentioned, this effect was confirmed
in experiment described  in the Rothe paper \cite{rothe}.

\section{The Breit--Wigner model}

In general the spectral density $\omega(E)$ has properties similar to the scattering amplitude, i.e., it can be
decomposed into a threshold factor, a pole-function $P(E)$ with a simple pole (often modeled by a Breit-Wigner)
and a smooth from factor $F(E)$.
So, we can write $\omega(E)= {\it\Theta}(E-E_{\rm min})\,(E-E_{\rm min})^{\alpha_{l}}\,P(E)\,F(E) $,
where
$\alpha_{l}$ depends on the angular momentum $l$ through $\alpha_{l} = \alpha + l$, \cite{fonda}
(see equation (6.1) in \cite{fonda}),  $0 \leq \alpha <1$)
 and ${\it\Theta}(E)$ is a step function: ${\it\Theta}(E) = 0\;\;{\rm  for}\;\; E \leq 0$
and ${\it\Theta}(E) = 1\;\;{\rm for}\;\;E>0  $. The simplest choice is to take $\alpha = 0, l=0, F(E) = 1$ and
to assume that $P(E)$ has a Breit--Wigner form.
It turns out that the decay curves obtained in this simplest case are very similar in form to the curves calculated for
the above described
more general $\omega (E)$,
(see \cite{nowakowski} and analysis in \cite{fonda}).
So to find the most typical properties of the decay curve it is sufficient to make the relevant calculations for  $\omega (E)$ modeled by the the Breit--Wigner
distribution of the energy density.

\begin{figure}[h!]
\begin{center}
\includegraphics[width=68mm]{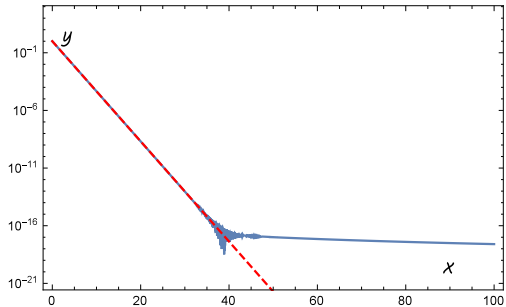}\\
\caption{Decay curves obtained for $\omega_{BW}(E)$ given by Eq. (\ref{omega-BW}).
Axes: $x =t / \tau_{0} $ --- time $t$ is measured in lifetimes
$\tau_{0}= \frac{\hbar}{{\it\Gamma}_{0}}$,    $y$ --- survival probabilities on a logarithmic scale (The solid line:  the decay curve ${\cal P}(t) = |a(t)|^{2}$; The dotted line:  the canonical decay curve
${\cal P}_{c}(t) = |a_{c}(t)|^{2}$. The case $s_{0} = \frac{E_{R}}{{\it\Gamma}_{0}} = 1000$.}
  \label{f1}
\end{center}
\end{figure}

The typical  form of the survival probability ${\cal P}(t)$ obtained in such a way is presented in Fig (\ref{f1}).
The calculations were made for $\omega (E)$ having the Breit--Wigner form
$\omega (E) \equiv \omega_{BW} (E)$,
\begin{equation}
\omega_{BW}(E) =  \frac{N}{2\pi}\,  {\it\Theta} (E - E_{min}) \
\frac{{\it\Gamma}_{0}}{(E-E_{0})^{2} +
(\frac{{\it\Gamma}_{0}}{2})^{2}}, \label{omega-BW}
\end{equation}
where $N$ is a normalization constant.

The case $\omega (E) = \omega_{BW}(E)$ is the typical case considered  in  numerous papers
and used therein  to model decay processes:
Among others, the Breit--Wigner model is often used to justify a belief  that there exists 
the "exponential time regime" of the decay process (see eg. \cite{peshkin,sluis} and Fig (\ref{f1})).
Therefore it is very important to analyze real form of the decay curves obtained using
$\omega (E) = \omega_{BW}(E)$ and this is why we consider this case in this paper. What is more,
substituting  $\omega_{BW}(E)$ into (\ref{a-spec-1}) allows one to find the analytical formula for the
amplitude $a(t)$. The result is  (see, eg. \cite{sluis,ku-2008,ku-2009})
\begin{eqnarray}
a(t) &=& N\,e^{\textstyle{- \frac{i}{\hbar} (E_{0} -
i\frac{{\it\Gamma}_{0}}{2})t}} \times \nonumber\\
&&\times \Big\{1 - \frac{i}{2\pi} \Big[
e^{\textstyle{\frac{{\it\Gamma}_{0}t}{\hbar}}}\,
E_{1}\Big(-\frac{i}{\hbar}(E_{R}
+ \frac{i}{2} {\it\Gamma}_{0})t\Big) \nonumber\\
&&\;\;\;\;\;+ (-1) E_{1}\Big(- \frac{i}{\hbar}(E_{R} -
\frac{i}{2} {\it\Gamma}_{0})t\Big)\,\Big]\, \Big\}, \label{a-E(1)}
\end{eqnarray}
where $E_{1}(x)$ denotes the integral--exponential function defined according to
\cite{abramowitz} and $E_{R} = E_{0} - E_{min}$.

The standard canonical form of the survival amplitude $a_{c}(t)$, is given by the following relation,
\begin{equation}
a_{c}(t) = \exp\,[{-i\frac{t}{\hbar}\,(E_{0} - \frac{i}{2}\,{\it\Gamma}_{0})}].
\end{equation}
${\it\Gamma}_{0}$ is the decay rate and $\frac{\hbar}{{\it\Gamma}_{0}} = \tau_{0}$ is the lifetime  (time $t$ and ${\it\Gamma}_{0}$
are measured in the rest reference frame of the particle).

It is convenient to consider the following function
\begin{equation}
\zeta (t) \stackrel{\rm def}{=} \frac{a(t)}{a_{c}(t)}. \label{zeta}
\end{equation}
There is $|\zeta (t)|^{2} = {\cal P}(t)/{\cal P}_{c}(t)$,
where ${\cal P}_{c}(t) = |a_{c}(t)|^{2}$ is the canonical exponential form of the decay law.
Analysis of properties of this function allows one to visualize all the more subtle differences between ${\cal P}(t)$ and
${\cal P}_{c}(t)$. For example, if one finds a time interval $[t_{1},t_{2}]$  such that  $\zeta (t) =const$ for $t \in [t_{1},t_{2}]$ this will mean that the survival probability
${\cal P}(t)$ has purely exponential form in this time interval.

\begin{figure}[h!]
\begin{center}
\includegraphics[width=68mm]{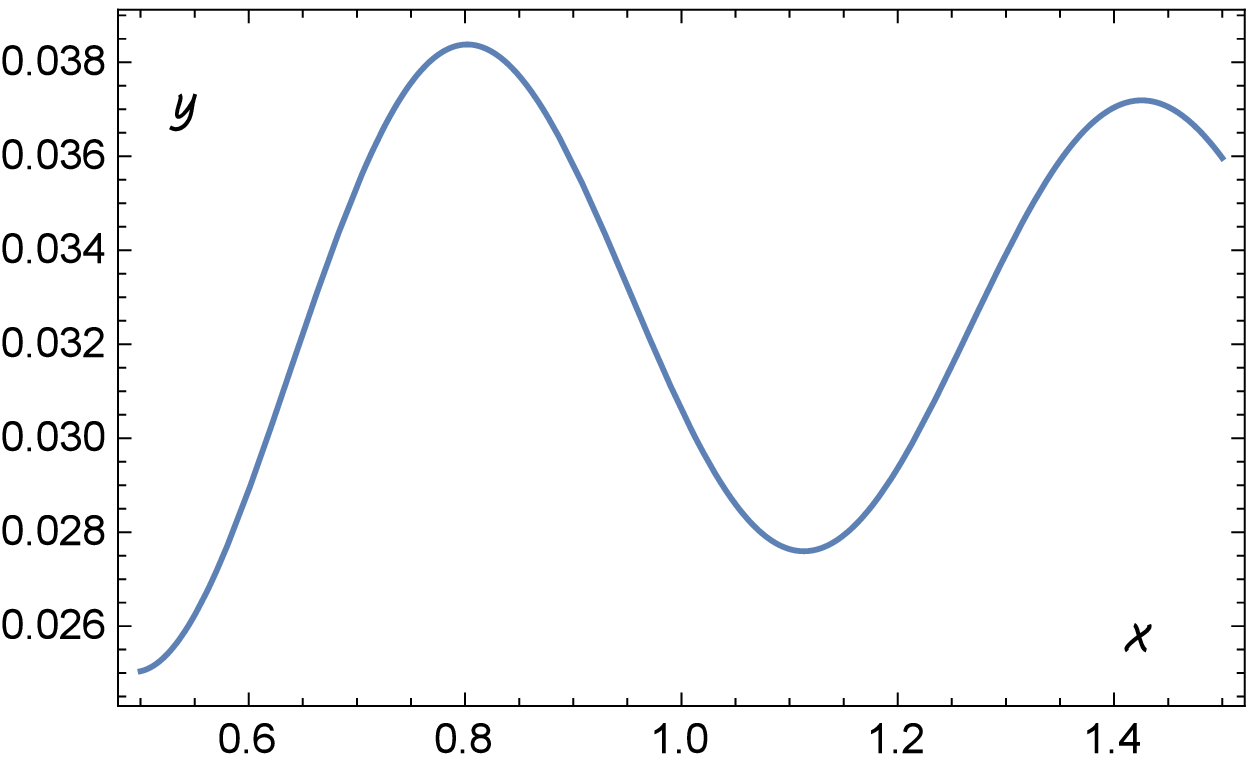}\\
\includegraphics[width=68mm]{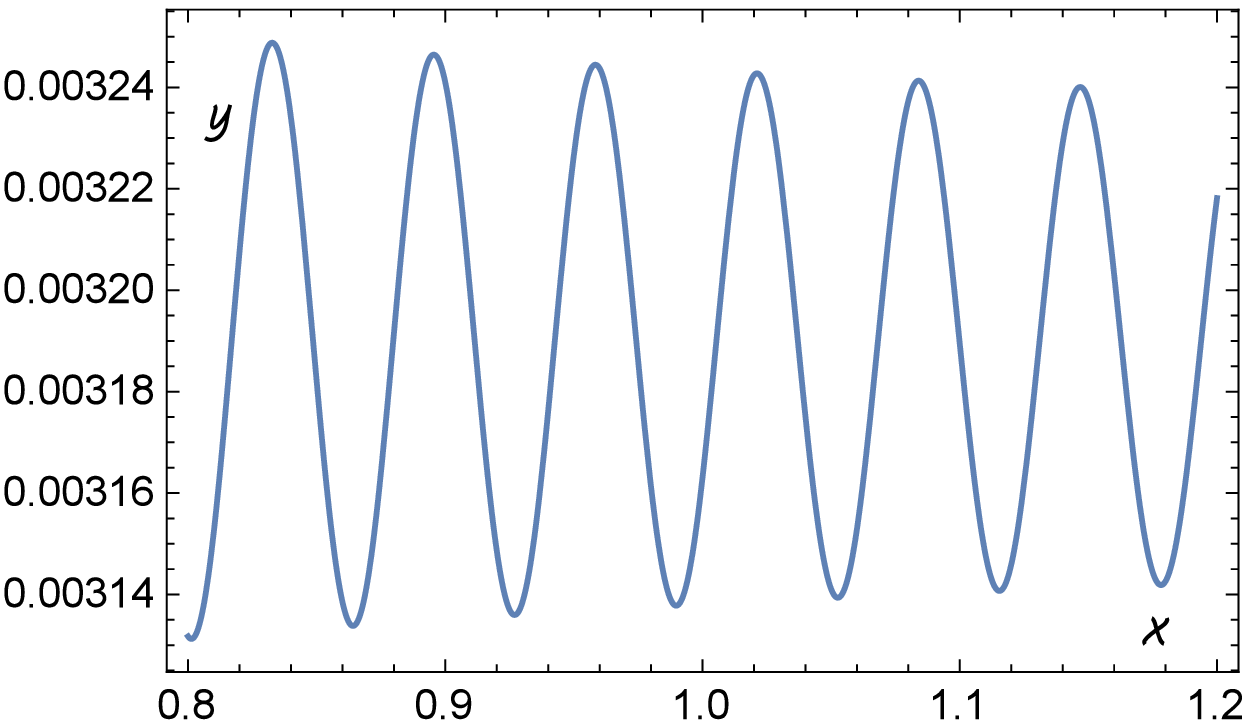}\\
\includegraphics[width=68mm]{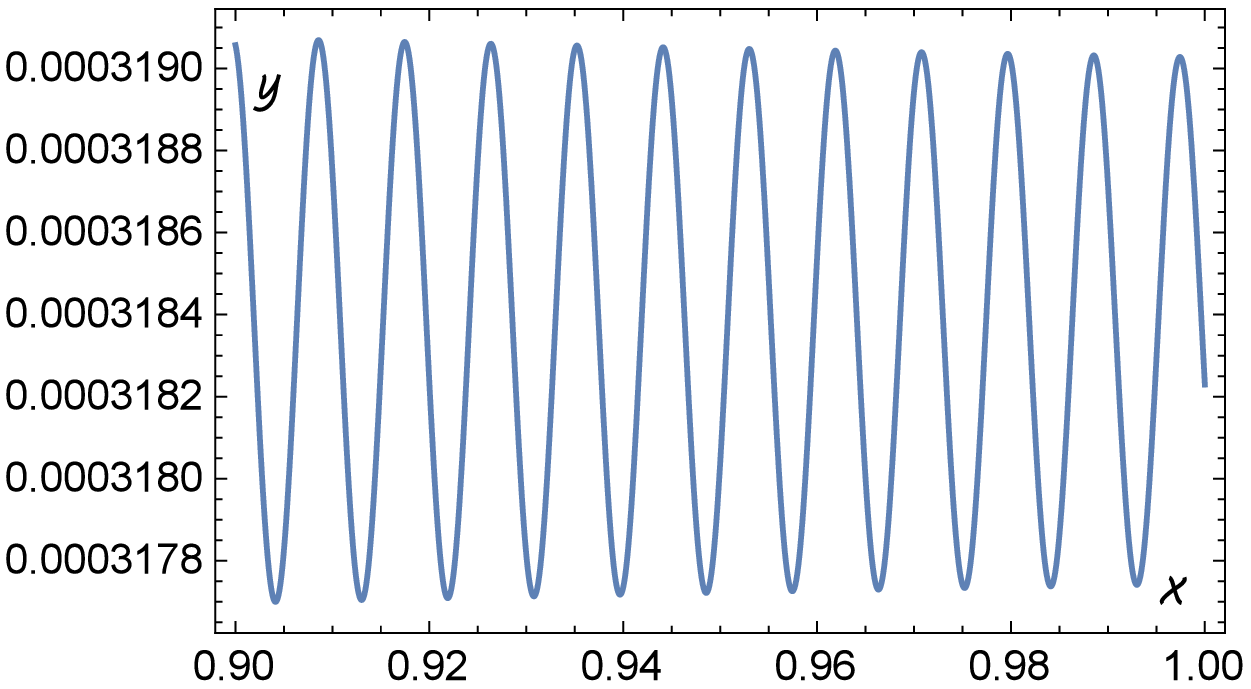}\\
\caption{A comparison of decay curves obtained for $\omega_{BW}(E)$ given by Eq. (\ref{omega-BW}) with canonical decay curves.
Axes: $x =t / \tau_{0} $ --- time $t$ is measured in lifetimes
$\tau_{0}$,   $y$ --- The function $f(t) = (|\zeta (t)|^{2} -1) = \frac{{\cal P}(t)}{{\cal P}_{c}(t)} - 1$, where $\zeta (t)$ is defined by the formula (\ref{zeta}).  The top panel:  $s_{R} = 10$. The middle panel: $s_{R} = 100$. The lower panel: $s_{R} = 1000$. }
  \label{f2}
\end{center}
\end{figure}

\begin{figure}[h!]
\begin{center}
\includegraphics[width=68mm]{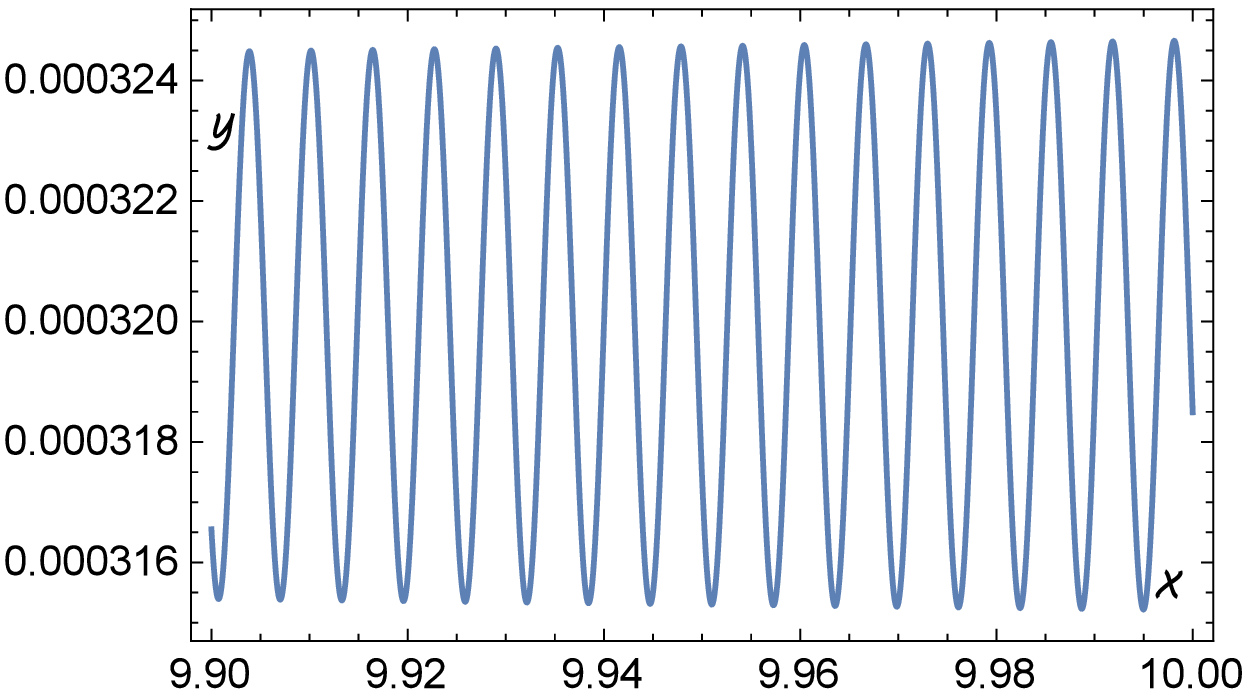}\\
\includegraphics[width=68mm]{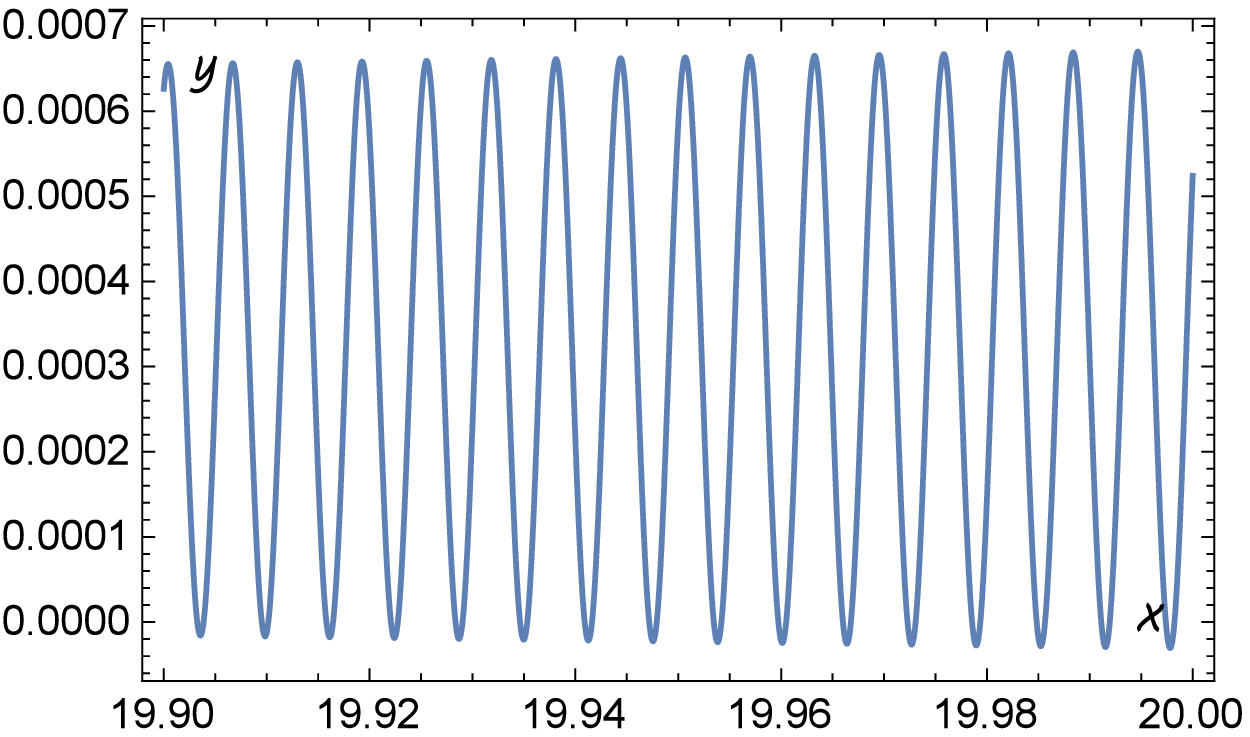}\\
\includegraphics[width=68mm]{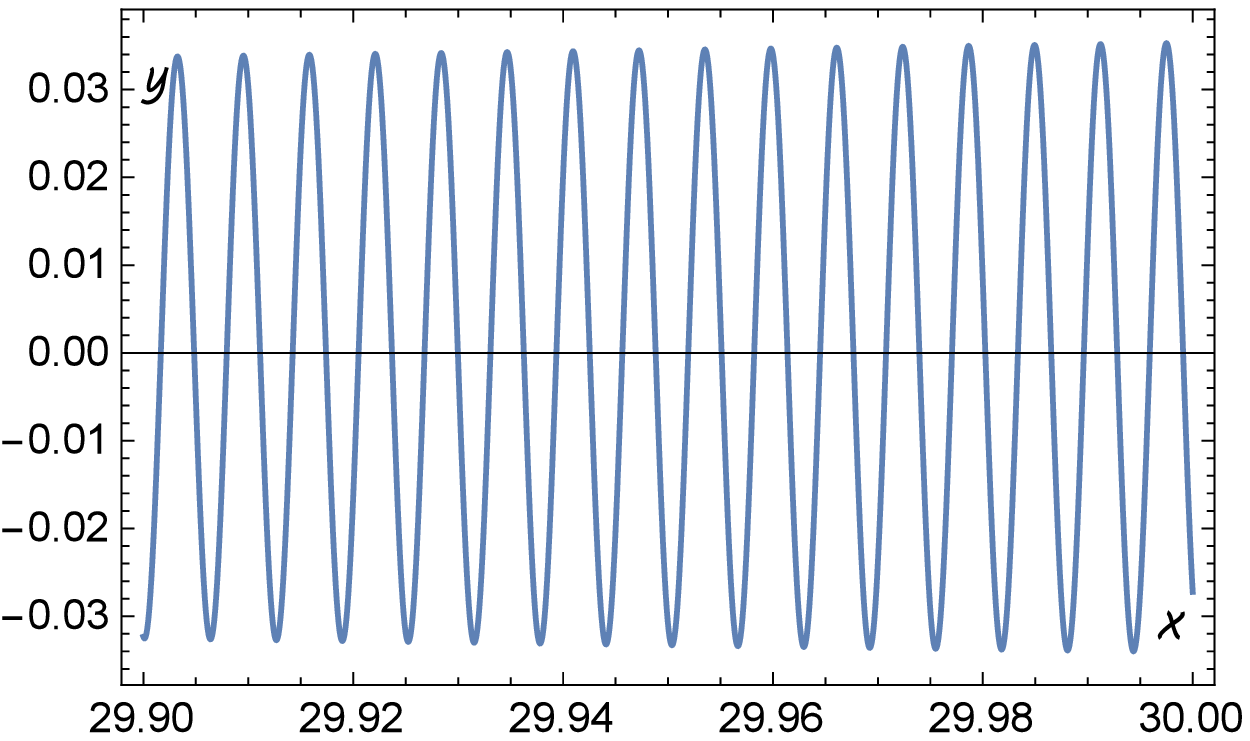}\\
\caption{A comparison of decay curves obtained for $\omega_{BW}(E)$ given by Eq. (\ref{omega-BW}) with canonical decay curves.
Axes: $x =t / \tau_{0} $ --- time $t$ is measured in lifetimes
$\tau_{0}$,   $y$ --- The function $f(t) = (|\zeta (t)|^{2} -1) = \frac{{\cal P}(t)}{{\cal P}_{c}(t)}\, - \,1$, where $\zeta (t)$ is defined by the formula (\ref{zeta}), ${\cal P}(t) = |a(t)|^{2}$,
${\cal P}_{c}(t) = |a_{c}(t)|^{2}$. The case  $s_{R} = 1000$.  }
  \label{f3}
\end{center}
\end{figure}

The function $\zeta(t)$ takes the following form in the case of the unstable system modeled by $\omega_{BW}(E)$:
\begin{eqnarray}
\zeta(t) &\equiv& N\; \Big\{1 - \frac{i}{2\pi} \Big[
e^{\textstyle{\frac{{\it\Gamma}_{0}t}{\hbar}}}\,
E_{1}\Big(-\frac{i}{\hbar}(E_{R}
+ \frac{i}{2} {\it\Gamma}_{0})t\Big) \nonumber\\
&&\;\;\;\;\;+ (-1) E_{1}\Big(- \frac{i}{\hbar}(E_{R} -
\frac{i}{2} {\it\Gamma}_{0})t\Big)\,\Big]\, \Big\}. \label{zeta1}
\end{eqnarray}
This function was used to find numerically $|\zeta (t)|^{2}$ for $\omega (E) = \omega_{BW}(E)$.
Results of numerical  calculations are presented
in Figs (\ref{f2}) and (\ref{f3}):
It turns out that in the case considered the form of  $|\zeta (t)|^{2}$ and ${\cal P}(t)$ depends on the ratio $s_{R} \stackrel{\rm def}{=} \frac{E_{R}}{{\it\Gamma}_{0}} \equiv \frac{E_{0} - E_{min}}{{\it\Gamma}_{0}}$.

The derivative of $\zeta (t)$ given by (\ref{zeta1}) equals \cite{ku-2008,ku-2009}
\begin{equation}
\frac{\partial \zeta (t)}{\partial t} = i\,\frac{N}{2\pi}\,\frac{{\it\Gamma}_{0}}{\hbar}\,e^{\textstyle{ \frac{{\it\Gamma}_{0}}{\hbar}t}}\,
E_{1}\Big(-\,\frac{i}{\hbar}(E_{R}+ \frac{i}{2}{\it\Gamma}_{0})t\Big). \label{d-zeta1}
\end{equation}
From the properties of the integral--exponential function $E_{1}(x)$ it follows that
the equation $\frac{\partial \zeta (t)}{\partial t} =0$ can be  satisfied at most for some isolated values of time  $t$.
So,
from the  formula (\ref{d-zeta1}) the conclusion follows: Within the model considered there is no time interval $[t_{1}, t_{2}]$, (where $t_{1} < t_{2}$),  in which $\zeta(t) = const$ for  $t \in [t_{1}, t_{2}]$, that is, there is no time interval in which
the survival probability ${\cal P}(t)$ has a pure exponential form.
This conclusion explains the results presented in Figs (\ref{f2}) and (\ref{f3}).

\section{The general case}
Results obtained for $\omega (E) = \omega_{BW}(E)$ and presented in Figs (\ref{f2}) and (\ref{f3}) can be understood as the interference of the pole contribution, $a_{pole}(t)$, into the survival probability $a(t)$ and the cut contribution, $a_{cut}(t)$, to $a(t) \equiv  a_{pole}(t) + a_{cut}(t)$. Analogous effects take place in all models of unstable states, in which unstable states are defined by poles of $\omega (E)$ in the complex plane.
Within the much more general Fock--Krylov theory \cite{fock} of unstable states the only condition that must be met by $\omega (E)$ is an absolute integrability of $\omega (E)$: If
$\omega (E)$ is an absolutely integrable function (even without the poles in the complex plane) then the Riemann--Lebesque lemma ensures that ${\cal P}(t) = |a(t)|^{2} \to 0$ as $t \to \infty$, where $a (t)$ is given by  (\ref{a-spec-1}). The question arises whether, in the most general case of $\omega (E)$ and $a(t)$ defined by  (\ref{a-spec-1}) (i. e. within the Fock--Krylov theory),
the effect described in the previous Section takes place or not.

So let us assume that the survival amplitude $a(t)$ is defined by (\ref{a(t)}) and can be represented as a Fourier transform (\ref{a-spec-1}) of some absolutely integrable $\omega (E)$. Now we can consider
the general case of $\zeta (t)$ defined using this general $a(t)$.
From the definition (\ref{zeta}) it follows that the equivalent form of $\zeta (t)$ is
\begin{equation}
\zeta (t) \equiv e^{\textstyle{+\,i\,\frac{t}{\hbar}\,(E_{0} - \frac{i}{2}\,{\it\Gamma}_{0})}}\;a(t). \label{zeta-1}
\end{equation}
Hence
\begin{eqnarray}
\frac{\partial \zeta (t)}{\partial t}
 &=& \frac{i}{\hbar}\, (E_{0} - \frac{i}{2}\,{\it\Gamma}_{0})\,\zeta (t) \;-\frac{i}{\hbar}\; h(t)\,\zeta (t), \label{d-zeta}
\end{eqnarray}
where
\begin{equation}
h(t) \stackrel{\rm def}{=} i\,\hbar\;\frac{1}{a(t)}\,\frac{\partial a(t)}{\partial t},\label{h(t)}
\end{equation}
is the effective Hamiltonian governing the time evolution in the subspace of unstable states ${\cal H}_{\parallel} = P {\cal H}$ and $P = |\phi\rangle \langle \phi|$
(see \cite{pra}  and also \cite{ku-2008,ku-2009} and references therein).
The subspace ${\cal H} \ominus {\cal H}_{\parallel} = Q {\cal H} = (\mathbb{I} - P) {\cal H}$ is the subspace of decay products.
The equivalent formula for $h(t)$ has the following form
\begin{equation}
h(t) \equiv \frac{\langle \phi|H|\phi (t) \rangle}{a(t)}. \label{h(t)-eq}
\end{equation}
If $\langle \phi |H| \phi\rangle $ exists then  using unitary evolution operator $U(t)$ and projection operators $P$ and $Q$ the last relation can be rewritten as follows
\begin{equation}
h(t) = \langle \phi|H|\phi \rangle  \;+\; \frac{\langle \phi|HQ\,U(t) |\phi \rangle}{a(t)}. \label{h(t)-1}
\end{equation}

Let us assume now that $\langle \phi|H|\phi \rangle$ exists and
there exists instants
$0< t_{1} < t_{2} < \infty $ of time $t$  such that for any $t \in (t_{1}, t_{2})$ there is
$\zeta (t) = \zeta (t_{1}) = \zeta (t_{2}) = const \, \stackrel{\rm def}{=} \,c_{\phi} \neq 0 $. In this case there should be $\frac{\partial \zeta (t)}{\partial t} = 0$ for all $t \in (t_{1},t_{2})$.
Taking into account that by definition $\zeta (t) \neq 0$  from (\ref{d-zeta}) we conclude that it is possible only and only if for $t_{1} \leq t \leq t_{2}$,
\begin{equation}
h(t) \, -\, (E_{0} - \frac{i}{2}\,{\it\Gamma}_{0}) = 0,\label{h-E}
\end{equation}
that is if and only if
\begin{equation}
h(t) = h(t_{1}) = h(t_{2}) = const \stackrel{\rm def}{=}  c_{h} \neq 0, \label{h=h1=h2}
\end{equation}
for $t_{1} \leq t \leq t_{2}$. Using  (\ref{h(t)-1}) and the property $|\phi (t) \rangle = U(t)\,|\phi\rangle$ one  concludes that the equality $h(t) = h(t_{2}) = c_{h}$ can take place if
\begin{equation}
\frac{\langle \phi|HQ\,U(t) |\phi \rangle}{a(t)} = \frac{\langle \phi|HQ\,U(t_{2}) |\phi \rangle}{a(t_{2})}. \label{h1=h2}
\end{equation}
Taking into account the group properties of the one--parameter family of unitary operators $U(t)$
we can use in (\ref{h1=h2}) the product $U(t)\,U(t_{2} - t)\equiv U(t_{2})$ instead of $U(t_{2})$. Next using the complex function $\lambda (t_{2},t) \stackrel{\rm def}{=} \frac{a(t_{2})}{a(t)}$ one can replace the relation (\ref{h1=h2})  by the following one
\begin{equation}
\langle \phi |HQ\,U(t)\,\Big[\,\lambda(t_{2},t)|\phi \rangle\;-\;U(t_{2} - t)|\phi\rangle\Big]=0. \label{h1=h2-a}
\end{equation}
This condition can be satisfied in two cases: The first one is
\begin{equation}
U(t_{2} - t)|\phi\rangle \,-\, \lambda(t_{2},t)|\phi \rangle \; =\;0, \label{d-zeta=0-1}
\end{equation}
and the second one occurs when $[\lambda(t_{2},t)|\phi \rangle\;-\;U(t_{2} - t)|\phi\rangle] \neq 0 $ and
vectors $(\langle \phi |H)^{+} = H|\phi\rangle$ and $ Q\,U(t)\,[\lambda(t_{2},t)|\phi \rangle\;-\;U(t_{2} - t)|\phi\rangle] $ are orthogonal to each other.

The first case means that $\frac{\partial \zeta (t)}{\partial t} = 0$ if and only
if the vector $|\phi \rangle$ representing an unstable state of the system is an eigenvector for the unitary evolution operator $U(t)$.  As we noted earlier
the evolution operator $U(t)$ and the total Hamiltonian $H$ of the system have common eigenvectors. This means that $\frac{\partial \zeta (t)}{\partial t} = 0$ for $t \in (t_{1},t_{2})$ if and only if the unstable state $|\phi\rangle$ of the system is an eigenvector for $H$, which is in contradiction with the property that the vector $|\phi\rangle$ representing the  unstable state cannot be the eigenvector for the total Hamiltonian $H$.

The second case: From the definition of the projectors $P$ and $Q$ it follows that this case can be realized only if the vector $H|\phi \rangle$ is proportional to the vector $|\phi\rangle$:  $H|\phi \rangle = \alpha_{\phi} |\phi\rangle$,
that is similarly to the first case $\frac{\partial \zeta (t)}{\partial t} = 0$ if and only
if the vector $|\phi\rangle$ representing the unstable state of the system considered is an eigenvector for the total Hamiltonian $H$,
which is again in clear contradiction with the condition that the vector $|\phi\rangle$ representing the  unstable state cannot be the eigenvector for the total Hamiltonian $H$.

Taking into account implications of the above two possible realizations of the relation (\ref{h1=h2-a})
we conclude that the supposition that such time interval $[t_{1},t_{2}]$ can exist that $\zeta (t) = const = \zeta(t_{1}) = \zeta(t_{2})$ for $t \in (t_{1},t_{2})$ is false. So taking into account the definition of $\zeta (t)$ the following conclusion follows:
Within the approach considered in this paper  for any time interval $[t_{1},t_{2}]$ the decay law can not be described by the exponential function of time. This conclusion is the general one. It does not depend on models of quantum unstable states and confirms the similar conclusion  drawn earlier for the Breit--Wigner model.

\section{Final remarks}
Summing up the oscillating decay curves of  one component unstable system can not be considered as something extraordinary or as anomaly: It seems to be a universal
feature of the decay process.
Oscillatory modulated decay curves are usually observed in two-- or more component unstable systems. A typical example of such systems is a neutral meson complex. From the results presented above it follows that such an effect can be also  observed in one component quantum unstable systems at the "exponential" decay regimes of times. What is more the oscillatory modulation of decay curves at the "exponential" decays regime takes place even in the quantum unstable system modeled by the Breit--Wigner distribution of the energy density.
In general,
 the oscillatory modulation of the survival probability at the "exponential" decay regime and thus the decay curves with model depending amplitude and oscillations period
takes place even in the case of one component unstable systems
modeled by  any physically acceptable form of $\omega (E)$.
From results of the model calculations presented in Figs (\ref{f2}) and (\ref{f3}) it follows that at the initial stage of the "exponential" (or "canonical") decay regime
the amplitude of these oscillations  may be much less than the accuracy of detectors. Then  with increasing time  the amplitude of oscillations grows (see Fig. (\ref{f3})), which increases the chances of observing them.
This is a true quantum picture of the decay process at the so--called "exponential" regime of times which should be taken into account when interpreting decay experiments
with one component unstable systems.


\begin{thebibliography}{19}
\bibitem{rutheford}
Ruthheford E.,
  {\em  Philosophical Magazine}, {\bf XLIX}, 1 and 161, (1900).
\bibitem{rutheford1}
Rutherford E. and F. Soddy F.,
    {\em Philosophical Magazine}, {\bf IV}, 370 and 569, (1902).
\bibitem{WW}
Weisskopf V. F. and Wigner E. T.,
    {\em Z. Phys.}, {\bf 63}, {54}, ({1930});
    {\bf 65}, 18, ({1930}).
\bibitem{khalfin}
Khalfin L. A.,
    {\em Zh. Eksp. Teor. Phys. (USSR)}, {\bf 33}, {1371}, ({1957}) [in Russian],
    [{\em Sov. Phys. --- JETP}, {\bf 6}, {1053}, ({1958})].
\bibitem{fonda}
Fonda L., Ghirardii G.C. and  Rimini A.,
    {\em Rep. on Prog. in Phys.} {\bf 41}, {587}, ({1978}).
\bibitem{norman}
E. B.  Norman, {\em et al},
    {\em Phys. Rev. Lett.}, {\bf 60},  {2246}, ({1988}).
\bibitem{rothe}
Rothe C., Hintschich S. I. and Monkman A. P.,
    {\em Phys. Rev. Lett.}, {\bf 96}, {163601}, ({2006}).
\bibitem{misra}
Misra B., Sudarshan E. C. G.,
    {\em Journal of Mathematical Physics}, {\bf 18}, {745}, ({1977}).
\bibitem{fischer}
Fischer M. C., {\em et al},
    {\em Phys. Rev. Lett.}, {\bf 87}, {040402}, ({2001}).
\bibitem{itano}
Itano W. M. {\em et al},
    {\em Phys. Rev.}, {\bf A 41}, {2295}, ({1990}).
\bibitem{patil}
Patil Y. S., {\em et al},
    {\em Phys. Rev. Lett.}, {\bf 115}, {140402}, ({2015}).
\bibitem{peshkin}
Peshkin M.,  Volya A. and  Zelevinsky V.
    {\em Europhysics Letters}, {\bf 107}, {40001}, ({2014}).
\bibitem{litwinov1}
Litvinov Yu. A., {\em et al}
    {\em Phys. Lett.}, {\bf B 664}, {162}, ({2008}).
\bibitem{kienle}
Kienle P., {\em et al},
    {\em Phys. Lett.}, {\bf B 726}, {638}, ({2013}).
\bibitem{fock}
Krylov N. S., Fock V. A.,
    {\em Zh. Teor. Eksp. Fiz.}, {\bf 17}, {93}, ({1947}) [in Russian];
    Fock V. A.,
    {\em Fundamentals of Quantum mechanics}
  (Mir Publishers, Moscow, 1978).
\bibitem{nowakowski}
Kelkar N. G, Nowakowski,
{\em J. Phys. A: Math. Theor.}, \textbf{43}, 385308,  (2010).
 \bibitem{giraldi}
 Giraldi F.,
    {\em Eur. Phys. J.}, {\bf D 69}, {5}, ({2015}).
\bibitem{giacosa2}
Giacosa F.,
    {\em Found. of Phys.}, {\bf 42}, {1262}, ({2012}).
\bibitem{goldberger}
Goldberger M. L., Watson K. M.,
    {\em Collision theory},
       (Wiley, 1964).
\bibitem{Paley}
Paley R. E. A. C.,
    {\em Fourier transforms in the complex domain},
    (American Mathematical Society, New York, 1934).
\bibitem{sluis}
Sluis K. M., Gislason E. A.,
    {\em Phys. Rev.}, {\bf A 43}, {4581}, ({1991}).
\bibitem{ku-2008}
Urbanowski K.,
    {\em Eur. Phys. J.}, {\bf C 58}, {151}, ({2008}).
 \bibitem{ku-2009}
 Urbanowski K.,
    {\em Cent. Eur. J. Phys.}, {\bf 7}, 696, (2009).
\bibitem{abramowitz}
    {\em Handbook of Mathematical Functions With Formulas, Graphs, and Mathematical Tables},
        Eds: Abramowitz Milton and Stegun Irene A.,
        (Dover Publications Inc., New York, 1964).
 \bibitem{pra}
 Urbanowski K.,
    {\em Phys. Rev.}, {\bf A 50}, {2847}, ({1994}).
\end{thebibliography}
\end{document}